# Application of blockchain technology for educational platform

Matija Šipek[1,2], Martin Žagar[1], Branko Mihaljević[1] and Nikola Drašković[1]

[1] Web and Mobile Computing Department, RIT Croatia, Zagreb, Croatia
[2] CS Computer Systems

{matija.sipek, martin.zagar, branko.mihaljevic, nikola.draskovic}@g.rit.edu, matija.sipek@cs.hr

**Abstract.** Nowadays, huge amounts of data are generated every second, and a quantity of that data can be defined as sensitive. Blockchain technology has private, secure, transparent and decentralized exchange of data as 'native'. It is adaptable and can be used in a wide range of Internet-based interactive systems in academic and industrial settings. The essential part of programmable distributed ledgers such as Ethereum, Polkadot, Cardano and other Web 3.0 technologies are smart contracts. Smart contracts are programs executed on the global blockchain, the code is public as well as all of the data managed within the transactions, thus creating a system that is reliable and cannot be cheated if designed properly. In this paper, in order to make the educational system more transparent and versatile we will describe an educational learning platform designed as a distributed system.

## 1 Introduction

### 1.1 Problem with Centralized systems

Areas associated with database systems have a wide range of industrial application; finance, academia, messaging services, healthcare etc. These systems are mainly deployed in a centralized manner, meaning a single entity has major decision-making and authority. So, it is possible that a trusted party goes rogue and seizes control of data governance.

Another problem with centralized services is the operational cost and user capacity; namely, with the industry 4.0 revolution and the ever-increasing number of user and devices traditional server-based systems will not be able to provide a continuous service. Internet of Things (IoT), Machine Learning, Artificial Intelligence (AI), Machine-to-Machine communication and others technologies will require immense resources and availability, and in this case centralized systems have scalability limitations as the future massive demand expansions of computing power [1]. Traditional database applications can have a decent level of security with additional security operations such as firewalls, however the centralized nature and the lack of control from the user side rises concern about real level of privacy. Still, in centralized systems there is always a single point of failure, regardless of backup databases. Also, it is known that companies with large amounts user data sell this data without users understanding what is happening, thus the implementation of blockchain technology as data security and privacy level is important to protect that data.

### 1.2 Smart contracts

A commonplace everyday mechanism in the life of humans are contracts; a piece of data and rules that uphold promises between two untrusting parties in a fair manner. The key concept of Smart Contracts was introduced in late 1990s by Nick Szabo with his initial idea to secure relationships over computer networks [2].

Smart contracts are programs executed on the global blockchain, the code is public as well as all of the data managed within the transactions, thus creating a system that is reliable and cannot be cheated if designed properly. Unlike traditional software, smart contracts are immutable, meaning once deployed, the code contained within a contract cannot be changed, unless if it is deployed as a new instance.

Likewise, another key concept is that they are deterministic, meaning that the outcome of execution of a smart contract is the same for every executor within the given context of the transaction. These rules and standards are creating a safe "middleman" ecosystem where manual involvement of a Trusted Third Party (TTP) between two parties is completely eliminated and automated, aside small transaction fees used to power the network's holders. Further, we are aiming to use the most economically and ecologically efficient system in order to reduce cost and the global carbon footprint; therefore, we also compared several consensus mechanisms such as Proof-of-Work (PoW), Proof-of-Stake (PoS) and Proof-of-Authority (PoA) [3].

Blockchain technology has private, secure, transparent and decentralized exchange of data as 'native'. It is adaptable and can be used in a wide range of Internet-based interactive systems in academic and industrial settings.

## 2 Methods and approach in system development

### 2.1 Simulation

Students were given the task of managing digital marketing activities for a fictitious smartphone brand that had a limited range of three phones with varying technical characteristics and a defined target demographic. The first round of the simulation is a test round, which allows students to become familiar with the interface after it has been set up. Various random events may occur during the simulation, such as the need for sales promotion support, a good cause, a distribution issue with the product, or a technical fault with the product, to which students must respond appropriately.

The Internet Marketing simulation was created as a three-step process: planning, execution, and reporting. The admin user, or professor in this example, has access to the backend controller system and can make changes to the rules before the simulation cycle begins. These rules are open to the public and do not affect game fairness because they are based on previous smart contract regulations.

### 2.2 Planning stage

The planning stage focuses on budgeting, keywords, and targeting possibilities. Each team (or, in this case, Blockchain node) must decide on money allocation and spending for each turn. Product, brand, and other communication channels and activities must all be divided within the weekly budget. Additionally, as is the case in reality, more targeted expenditure (e.g., limited to one product and/or fewer channels/activities) is more efficient. This stage was entirely built using JavaScript technology, and it can be accessed from the application's front-end.

### 2.3 Execution of the plan

Students can make some minor adjustments to the plan's implementation to improve the advertising effort. The keyword (or numerous keywords, depending on the channel) must be chosen by each student or team. Depending on the style of class delivery, this phase is done on a daily or weekly basis. Students can clearly identify their target audience in order to optimize the total promotional effort (s). Targeting is a platform-specific term that applies to promotional efforts. For success, the right mix of messaging strategy and targeting is critical.

### 2.4 Reporting stage

Students receive official reports (digital platform insights/analytics) and some less formal reports once the round/turn is completed (sales force feedback). Students will have to pay a set amount of money for the market report (which will include statistics on the competition and customer preferences). This backend process is built using distributed Blockchain technologies instead of a single centralized database, and we've developed Blockchain-based reporting.

## 3 Technology

One of the key goals was to focus on decentralization and differential privacy as privacy assurances for both students and academics. The system is divided into three primary sectors, each of which has specific responsibilities and is linked by an intermediary level that manipulates and shares data.

In comparison to the EVM (Ethereum Virtual Machine) cost of transactions, we created a functioning prototype that employs the underlying technologies that allow free transactions per se [8]. Nonetheless, by simply changing the target network and altering some communicational variables, the system can be transferred to any EVM-based blockchain network.

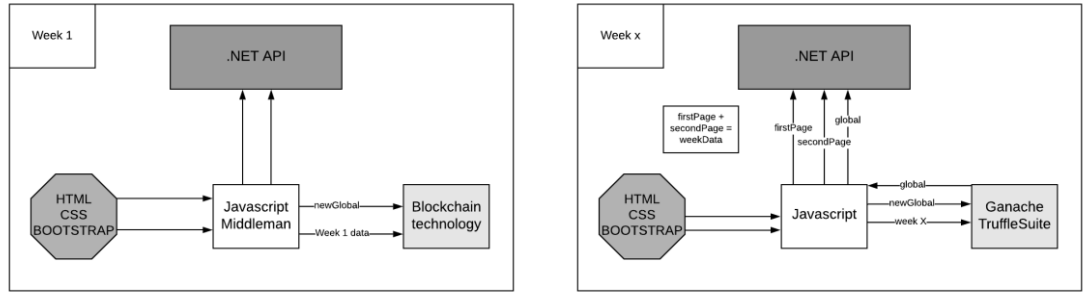

**Figure 1 Key communication exchange**

As it can be seen from the figure one, our system comprises of three interconnected independent parts; the first is the frontend of the application which handles all user communication, the second part is .NET API which is responsible for all heavy-weight computational efforts, and finally we have the blockchain part where crucial data exchange happen.

### 3.1 Truffle Suite

Truffle Suite is a development environment that includes all of the tools needed to complete the whole development process of a blockchain platform. Truffle handles contract artifact management and provides support for easy deployments, library linking, and complicated Ethereum applications. It also allows you to write automated tests in both JavaScript and Solidity for smart your contracts. From dependencies in our deployment, we were using Truffle v5.0.2 (core: v5.0.2) and truffle-contract v3.0.6.

### 3.2 Ganache

Ganache is a local in-memory Ethereum blockchain environment that mimics the behavior of the distributed ledgers in the real world. The system displays the current state of all accounts, including their addresses, private keys, transactions, balances, and the log output of Ganache's internal blockchain, which includes responses and other important debugging data. Furthermore, ganache has additional mining features, including the ability to modify block times to best suit your development needs and examine all blocks and transactions to learn more about what's going on behind the scenes. For development we have used Ganache v2.4.0 with JSON Remote Procedure Call (RPC) server setup, the price of each unit of gas is set up at 20000000000 Wei, the gas limit is 6721975 units and the hard fork is set to Petersburg.

*3.3    Web3.js*

Web3 is a collection of protocols aimed at making the internet more decentralized, verifiable, and secure. The web3.js library is a set of modules that allow Ethereum ecosystem functionalities to be connected to traditional applications. We also implemented web3.js v1.3.0 as a JavaScript web provider to link Truffle Suite and Ganache to the browser.

*3.4    MetaMask*

MetaMask provides a key vault, secure login, token wallet, and token exchange as a browser extension and a mobile app, giving you everything you need to manage your digital assets. MetaMask is a browser plugin that acts as an Ethereum wallet, enabling websites to request Ethereum accounts and thereby run Ethereum decentralized applications (dApps). MetaMask does this by including a provider object that identifies an Ethereum user, resulting in an API that can read data from the blockchains. In our system, we have used the MetaMask v4.0.2.

*3.5    Solidity*

Solidity is an object-oriented programming language with a statically typed syntax used for smart contract development on a variety of blockchain systems, the most prominent of which is Ethereum. We used solidity v0.5.0 for our system.

*3.6    Additional technologies*

The API handles the most data and is the primary method for sales force calculations. JavaScript calls are performed to the API via specific controllers during the simulation, which may handle user interaction with or without blockchain to do initial computations and conclude the weekly turn report.

For the .NET Core version, the API business logic calculation subsystem uses C# with Microsoft (R) Build Engine version 16.2.32702+c4012a063 and the project SDK Microsoft.NET.Sdk.Web with the netcoreapp2.2 target framework.

## 4    Conclusion

One of the key objectives of our system was to place a strong emphasis on privacy guarantees for both students and teachers, such as decentralization and differential privacy. The instructor-accessible admin tools, as well as the user panel and other settings based on distributed Blockchain reporting, allow for a high level of customization in our program. Furthermore, our application may function in multi-user mode, allowing students (or teams of students) to compete against one another.

# 5 The References Section